\documentclass[copyright,creativecommons]{eptcs}
\providecommand{\event}{FBTC 2010} 
\providecommand{\volume}{??}
\providecommand{\anno}{20??}
\providecommand{\firstpage}{1}
\providecommand{\eid}{??}
\usepackage{breakurl}        
\usepackage{amsmath}
\usepackage{amsthm}
\usepackage{amsfonts}

\usepackage{amssymb}
\usepackage{amsbsy}
\usepackage{prooftree}
\usepackage{graphicx}
\usepackage{mathptmx}
\usepackage{eurosym}
\usepackage{txfonts}
\usepackage{listings}
\usepackage{wasysym,xspace}
\newtheorem{notation}{Notation}
\newtheorem{definition}{Definition}
\newtheorem{theorem}{Theorem}
\title{Types for BioAmbients\thanks{This research is funded by the BioBITs Project (\emph{Converging
Technologies} 2007, area: Biotechnology-–ICT), Regione Piemonte.}}
\author{Sara Capecchi \hspace{0.5cm} Angelo Troina
\institute{Dipartimento di Informatica, Universit\`a di Torino}
\email{\{capecchi,troina\}@di.unito.it}
}
\def\titlerunning{Types for BioAmbients}
\def\authorrunning{S. Capecchi, A. Troina}

\newcommand{\act}{\ensuremath{\pi}}
\newcommand{\dir}{\ensuremath{\$}}
\newcommand{\deff}{\ensuremath{::=}}
\newcommand{\sep}{\ensuremath{~\mathbf{|}~ }}
\newcommand{\capb}{\ensuremath{M}}
\newcommand{\nul}{\ensuremath{\mathbf{0}}}

\newcommand{\bname}{\ensuremath{b}}
\newcommand{\m}{\ensuremath{m}}
\newcommand{\tname}{\ensuremath{t}}
\newcommand{\h}{\ensuremath{h}}
\newcommand{\n}{\ensuremath{n}}

\newcommand{\PP}{\ensuremath{P}}

\newcommand{\R}{\ensuremath{R}}
\newcommand{\SP}{\ensuremath{S}}
\newcommand{\G}{\ensuremath{G}}
\newcommand{\GE}[1]{\ensuremath{\G_{#1}}}
\newcommand{\T}{\ensuremath{T}}
\newcommand{\Gbar}{\ensuremath{\overline{G}}}
\newcommand{\Q}{\ensuremath{Q}}
\newcommand{\aname}{\ensuremath{a}}
\newcommand{\cname}{\ensuremath{c}}

\newcommand{\SGn}{\ensuremath{\mathcal{S}}}
\newcommand{\CGn}{\ensuremath{\mathcal{C}}}

\newcommand{\GR}{\ensuremath{(\SGn,\CGn)}}

\newcommand{\SG}{\ensuremath{\SGn(\G)}}
\newcommand{\CG}{\ensuremath{\CGn(\G)}}

\newcommand{\SGe}[1]{\ensuremath{\SGn(#1)}}
\newcommand{\CGe}[1]{\ensuremath{\CGn(#1)}}

\newcommand{\loc}{\ensuremath{\mathsf{loc}}}
\newcommand{\ptc}{\ensuremath{\mathsf{ptc}}}
\newcommand{\ctp}{\ensuremath{\mathsf{ctp}}}
\newcommand{\sts}{\ensuremath{\mathsf{sts}}}

\newcommand{\entern}{\ensuremath{\mathsf{enter} }}
\newcommand{\acceptn}{\ensuremath{\mathsf{accept} }}
\newcommand{\exitn}{\ensuremath{ \mathsf{exit} }}
\newcommand{\expeln}{\ensuremath{ \mathsf{expel} }}
\newcommand{\mergepn}{\ensuremath{ \mathsf{merge\oplus} }}
\newcommand{\mergemn}{\ensuremath{\mathsf{merge-} }}

\newcommand{\enter}[1]{\ensuremath{\entern \ #1}}
\newcommand{\accept}[1]{\ensuremath{\acceptn \ #1}}
\newcommand{\exit}[1]{\ensuremath{ \exitn \ #1}}
\newcommand{\expel}[1]{\ensuremath{ \expeln \ #1}}
\newcommand{\mergep}[1]{\ensuremath{ \mergepn \ #1}}
\newcommand{\mergem}[1]{\ensuremath{\mergemn \ #1}}

\newcommand{\inG}[2]{\ensuremath{\dir #1?\{#2\}}}
\newcommand{\outG}[2]{\ensuremath{\dir #1!\{#2\}}}
\newcommand{\inE}[3]{\ensuremath{#1 \ #2?\{#3\}}}
\newcommand{\outE}[3]{\ensuremath{#1 \ #2!\{#3\}}}

\newcommand{\restr}[2]{\ensuremath{(\nu \ #1)#2}}
\newcommand{\pa}{\ensuremath{\ | \ }}
\newcommand{\repl}[1]{\ensuremath{ ! #1}}
\newcommand{\amb}[2]{\ensuremath{#1 [\![#2]\!]}}
\newcommand{\comm}[1]{\ensuremath{\act.#1}}
\newcommand{\capP}[1]{\ensuremath{\st.#1}}
\newcommand{\comch}[1]{\ensuremath{\sum_{i \in I}\act_i.#1_i}}
\newcommand{\capch}[1]{\ensuremath{\sum_{i \in I}\capb_i.#1_i}}

\newcommand{\gt}{\ensuremath{\G}}
\newcommand{\st}{\ensuremath{S}}
\newcommand{\ga}{\ensuremath{\Gamma}}
\newcommand{\de}{\ensuremath{\Delta}}

\newcommand{\typer}[2]{\ensuremath{\ga \vdash #1 :#2 }}
\newcommand{\typd}[2]{\ensuremath{\ga \vdash #1 :#2 }}
\newcommand{\typdG}[2]{\ensuremath{\ga,#1 :#2 \vdash #1 :#2 }}
\newcommand{\typerG}[3]{\ensuremath{#1 \vdash #2 :#3 \triangleright \de}}
\newcommand{\typerD}[2]{\ensuremath{\ga \vdash #1 :#2 \triangleright \de}}
\newcommand{\typerDe}[3]{\ensuremath{\ga \vdash #1 :#2 \triangleright #3}}

\newcommand{\comp}[2]{\ensuremath{ #1 \asymp #2 }}

\newcommand{\trule}[1]{\ensuremath{\lfloor\text{\sc{#1}}\rfloor}}

\newcommand{\subst}[3]{\ensuremath{#1[#2 \leftarrow #3] }}

\newcommand{\ct}{\ensuremath{\gamma}}
\newcommand{\ch}[1]{\ensuremath{ch\{#1\}}}
\newcommand{\sy}{\ensuremath{s}}

\newcommand{\sync}{\ensuremath{(\Gbar_1, \Gbar_2)^\lab}}
\newcommand{\lab}{\ensuremath{\ell}}
\newcommand{\ea}{\ensuremath{\mathsf{\entern/\acceptn}}}
\newcommand{\ee}{\ensuremath{\mathsf{\exitn/\expeln}}}
\newcommand{\mm}{\ensuremath{\mathsf{\mergepn/\mergemn}}}

\newcommand{\red}[2]{\ensuremath{#1 \longrightarrow #2}}
\newcommand{\congr}[2]{\ensuremath{#1 \equiv #2}}

\newcommand{\warnname}{\ensuremath{W}}
\newcommand{\warn}[1]{\ensuremath{\warnname(#1)}}
\newcommand{\exerror}[2]{\ensuremath{\mathsf{expelError} \ \ \amb{#1}{#2}}}
\newcommand{\merror}[2]{\ensuremath{\mathsf{mergeError} \ \ \amb{#1}{#2}}}

\newcommand{\stc}{\ensuremath{\equiv}}
\newcommand{\error}[2]{\ensuremath{\mathsf{Error} \ \ \amb{#1}{#2}}}

\begin{document}
\maketitle
\begin{abstract}
The BioAmbients calculus is a process algebra suitable for representing compartmentalization, molecular localization and movements between compartments.
In this paper we enrich this calculus with a static type system classifying each ambient with group types specifying the kind of compartments in which the ambient can stay. The type system ensures that, in a well-typed process, ambients cannot be nested in a way that violates the type hierarchy. Exploiting the information given by the group types, we also extend the operational semantics of BioAmbients with rules signalling errors that may derive from undesired ambients' moves (i.e. merging incompatible tissues). Thus, the signal of errors can help the modeller to detect and locate unwanted situations that may arise in a biological system, and give practical hints on how to avoid the undesired behaviour.
\end{abstract}

\section{Introduction}

BioAmbients~\cite{BA} is a variant of the Ambient Calculus~\cite{CG98}, in which compartments are described as a hierarchy of boundary ambients. This hierarchy can be modified by suitable operations that have an
immediate biological interpretation; for example, the interactions between compounds that reside in the cytosol and in the nucleus of a cell could be modelled via parent--child communications. Thus, BioAmbients is quite suitable for the representation of various aspects of molecular localization and compartmentalization, such as the movement of molecules between compartments, the dynamic rearrangement that occurs between cellular compartments, and the interaction between the molecules in a compartmentalized context.

A stochastic semantics for BioAmbients is given in~\cite{BDP07}, and an abstract machine for this semantics is developed in~\cite{Phil09}. In~\cite{FQY05} BioAmbients is extended with an operator modelling chain-like biomolecular structures and applied within a DNA transcription example. In~\cite{PNN08} a technique for pathway analysis is defined in terms of static control flow analysis. The authors then apply their technique to model and investigate an endocytic pathway that facilitates the process of receptor mediated endocytosis.

In this paper we extend the BioAmbients calculus with a static type system that classifies each ambient with a group type $G$ specifying the kind of compartments in which the ambient can stay \cite{Cardelli2002160}. In other words, a group type $G$ describes the properties of all the ambients and processes of that group. Group types are defined as pairs \GR, where \SGn\ and \CGn\ are sets of group types. Intuitively, given $G=$\GR, \SGn\ denotes the set of ambient groups where ambients of type $G$ can stay, while \CGn\ is the set of ambient groups that can be crossed by ambients of type $G$. On the one hand, the set \SGn\ can be used to list all the elements that are allowed within a compartment (complementary, all the elements which are not allowed, i.e. repelled). On the other hand, the set \CGn\ lists all the elements that can cross an ambient, thus modelling permeability properties of a compartment.

Starting from group types as bases, we define a type system ensuring that, in a well-typed process, ambients cannot be nested in a way that violates the group hierarchy. Then, we extend the operational semantics of BioAmbients, exploiting the information given by the group types, with rules rising warnings and signalling errors that may derive from undesired compartment interactions. For example, while correctness of the \emph{enter}/\emph{accept} capabilities (that are used to move a compartment to the inside of another compartment) can be checked statically, the \emph{merge} capability (which merges two compartments into one) and the \emph{exit}/\emph{expel} capabilities (which are used to move a compartment from the inside to the outside of another compartment) could cause the movement of an ambient of type $G$ within an ambient of type $G'$ which does not accept it. In these cases, for example when incompatible tissues come in contact, an error signal is raised dynamically and the execution of the system is blocked. The modeller can exploit these signals as helpful \emph{debugging} information in order to detect and locate the unwanted situations that may arise in a biological system. Intuitively, they give practical hints on how to avoid the undesired behaviour.

In the last few years there has been a growing interest on the use of type disciplines to enforce biological properties. In~\cite{ADT08} a type system has been defined to ensure the wellformedness of links between protein sites within the Linked Calculus of Looping Sequences (see~\cite{BMM06}). In~\cite{FS08} three type systems are defined for the Biochemical Abstract Machine, BIOCHAM (see~\cite{BioCHAM}). The first one is used to infer the functions of proteins in a reaction model, the second one to infer activation and inhibition effects of proteins, and the last one to infer the topology of compartments. In~\cite{DGT09a} we have defined a type system for the Calculus of Looping Sequences (see~\cite{BMMT06}) to guarantee the soundness of reduction rules with respect to the requirement of certain elements, and the repellency of others. Finally, in~\cite{DGT09b} we have proposed a type system for the Stochastic Calculus of Looping sequences (see~\cite{BMMTT08}) that allows for a quantitative analysis and models how the presence of catalysers (or inibitors) can modify the speed of reactions.

\subsection{Summary}
The remainder of the paper is organised as follows. In Section~\ref{sec:syntax} we recall the original BioAmbients' syntax. In Section~\ref{sec:TypeSystem} we define our type system and in Section~\ref{sec:TypedOpSem} we give our typed operational semantics. In Section~\ref{sec:Examples} me formulate two motivating examples, namely we use our type system to analyse blood transfusions (rising errors in the case incompatible blood types get mixed) and spore protection against bacteriophage viruses. Finally, in Section~\ref{sec:Conc} we draw our conclusions.

\section{BioAmbients: Syntax}
\label{sec:syntax}

In this section we recall the BioAmbients calculus.  \emph{Ambients} represent bounded mobile entities that can be nested forming hierarchies. They provide an intuitive mean to model both \emph{membrane-bound compartments}, where the components of a compartment are isolated from the external environment, and \emph{molecular compartments} i.e. multi molecular complexes in which molecules can be partially isolated from the environment.
\emph{Capabilities} are  used to model  movements changing ambients hierarchies: they can be employed to model membranes fusion, molecules movement, complexes formation.
Finally, \emph{communications} model interactions between components within or across ambients boundaries.

The syntax is defined in Figure~\ref{tab:Syntax} and is the same as that of \cite{BA}. The only difference is that in our syntax ambients names are not optional. We give a name to each ambient in order to associate its type to it.
\emph{Ambient names} are ranged over by $\aname, \aname_1, \bname   \ldots$, \emph{channel names} are ranged over by $\cname, \cname_1, \ldots$, \emph{capability names} are ranged over by $\h, \h_1, \ldots$. We use $\n, \m$ to range over unspecified names and  \PP,\Q, \R, \T\ to range over processes.

\emph{Capabilities} syncronise using names (\h) and allow an ambient  (1) to enter in a sibling ambient accepting it (\enter{\h} /\accept{\h}), (2) to leave the parent ambient ( \exit{\h} /\expel{\h}), (3) to merge with a sibling forming a unique ambient (\mergep{\h} /\mergem{\h}).
\emph{Communications} on channels (\outG{\cname}{\m},\inG{\cname}{\m}) are prefixed by \emph{directions} (\dir) denoting different kinds of communications: local communications (\loc)  within the same ambient, sibling communications (\sts)  between sibling ambients, parent/child  (\ptc,\ctp)  between nested ambients.
Concerning processes syntax: inaction \nul\ is a special case of summation ($I = \emptyset$) and denotes the process doing nothing; restriction \restr{\n}{\PP} restricts the scope of the name \n \ to \PP; \PP\pa\Q\ denotes the parallel composition of \PP\ and \Q; \repl{\PP} stands for process replication; \amb{\aname}{\PP} describes a process \PP\ confined in an ambient named \aname; communication and capability choices (\comch{\PP}, \capch{\PP}) generalise communication and capability prefixes respectively ( \comm{\PP}, \capP{\PP} ) and represent standard choices.

\begin{figure}
	\centering
		\begin{tabular}{rclr}

\act & \deff &  &   {Actions}\\
     & \sep & \outG{\cname}{\m} &    {Output}\\
     & \sep & \inG{\cname}{\m} &    {Input}\\
     \\
\dir & \deff  &  &   {Directions}\\
     & \sep & \loc & {Intra-Ambient}\\
     & \sep & \sts & {Inter-siblings}\\
     & \sep & \ptc & {Parent to child}\\
     & \sep & \ctp & {Child to parent}\\
     \\
\capb & \deff  &    &   {Capabilities Prefixes} \\
     & \sep & \enter{}& {Entry}\\
     & \sep & \accept{}& {Accept}\\
     & \sep & \exit{}& {Exit}\\
     & \sep & \expel{}& {Expel}\\
     & \sep & \mergep{}&{Merge with} \\
     & \sep & \mergem{}&{Merge into} \\
     \\
\st  & \deff  &   \capb \h &   {Capabilities} \\
\\
\PP & \deff  &  &   {Processes}\\
& \sep & \nul& {Empty process}\\
& \sep & \restr{\n}{\PP}& {Restriction}\\
& \sep & \PP\pa\Q& {Composition}\\
& \sep & \repl{\PP}& {Replication}\\
& \sep & \amb{\aname}{\PP}& {Ambient}\\
& \sep & \comm{\PP}& {Communication prefix}\\
& \sep & \capP{\PP}& {Capability prefix}\\
& \sep & \comch{\PP}& {Communication choice}\\
& \sep & \capch{\PP}& {Capability choice}\\

\\
\end{tabular}
	\caption{BioAmbients: Syntax.\label{tab:Syntax}}
	
\end{figure}

\section{The Type System}\label{sec:TypeSystem}
We classify ambients names with \emph{group types} as in \cite{Cardelli2002160,m3}. Intuitively, the type \G\ of an ambient denotes the set of ambients where that particular ambient can stay:
   it describes, in terms of other group types
(possibly including \G),  the properties of all the
ambients and processes of that group.

Group types consist of two components and
are of the form \GR, where \SGn\ and \CGn\ are sets of group types.
The intuitive meanings of the types' sets are the following:
\begin{itemize}
	\item \SGn\ is the set of ambient groups where the ambients of group \G\ can stay;
\item \CGn\ is the set of ambient groups that \G-ambients can cross, i.e., those that
they may be driven into or out of, respectively, by \entern\ and \exitn\ capabilities.
\end{itemize}
Clearly for all \G\, \CG $\subseteq \SG$. If \G = \GR\ is a group type, we write \SG\ and \CG\
respectively to denote the components \SGn\ and \CGn\ of \G.
We call \GE{Univ} the type of  universal environments where each ambient can stay in.
Types syntax is given in Figure~\ref{tab:typesyntax}.

Besides group types we have:
\begin{itemize}
	\item Capability types: \sync\ is the type associated to a name \h\ through which  ambients of types $\Gbar_1$ and  $\Gbar_2$ can perform the  movements described by \lab.

	\item Channel types \ct: the types of the channels arguments which can be groups (\G) , capabilities  (\sy)   or channels (\ct).
\end{itemize}

\begin{notation}
Let \capb\ be a capability prefix and \sy=$(\Gbar_1, \Gbar_2)^{(\capb_1,\capb_2)}$ be a capability type, we say $\capb \in \sy$ if either $\capb = \capb_1$ or $\capb = \capb_2$.
\end{notation}
\noindent
We now define well-formedness for capability types.

\begin{definition}[\sy-Well-formedness]
\label{def:wf}
A capability type  \sync\ is well formed iff none of the following holds:
\begin{enumerate}
	\item  \lab= \ea\ and $\exists \G_i \in \Gbar_2,\ \G_j \in \Gbar_1:  \G_i\notin \CGe{\G_j}$
	\item   \lab= \ee\  and $\exists \G_i \in \Gbar_2,\ \G_j \in \Gbar_1:  \G_i\notin \CGe{\G_j}$
\end{enumerate}
\end{definition}
\noindent
Intuitively, a capability type \sync\ describing the entrance(exit) of an ambient of type $\G_j \in \Gbar_1$ into(out of) an ambient of type $\G_i \in \G_2$ is not correct if $\G_j$ cannot (cross)stay in $\G_i$.

\begin{figure}[t]
	\centering
\begin{tabular}{lclr}
$\gt_1 \ldots \gt_n $ &  &  & Group types
\\
\tname & \deff & \gt \sep \sy \sep \ct & Channels arguments \\
\sy & \deff & \sync & Capability types\\
\lab & \deff & \ea \ \sep \ \ee \ \sep \ \mm & Labels\\

\ct & \deff & \ch{\tname}  & Channels \\
\\
\\
\end{tabular}
\caption{Type syntax. \label{tab:typesyntax}}
\end{figure}	

%
\noindent
We now define the environment \ga\ mapping names to types:
\[
\ga ::= \emptyset \ | \  \ga,\m : \tname
\]
we assume that we can write \ga, \m : \tname\ only if \m\ does not occur in \ga, i.e. $\m \notin Dom(\ga)$
($Dom(\ga)$ denotes the domain of \ga, i.e., the set of names  occurring in \ga).
An environment \ga\ is well formed if  for each  name the associated type
 is well formed.

In the following we define compatibility between a group type and an argument type.

\begin{definition}[\sy-\G\ Compatibility]
\label{def:comp}
Given a capability type \sy=\sync\  and a group type \G\ we define their compatibility as follows:
\begin{center}
\comp{\sy}{\G}  iff  \sy\ is well formed and  at least one between $\G \in\Gbar_1$  and $\G \in\Gbar_2$ holds.
	
\end{center}
\end{definition}
\noindent
We can check the safety of BioAmbients processes using the rules in Figure~\ref{tab:Typing}.
Let \ga\ be type environment from which  we  derive the type of  names (rule \trule{Name});
typing rules for processes have the shape :
\[
	\typerD{\PP}{\Gbar}
\]
 where  \PP\ is a process, \Gbar\ is a set of group types representing  the types of the ambients in \PP\ and \de\ is a set of capability types collecting the capabilities in \PP.

 \trule{Inact} derives any group type \G\  for  the empty process (indeed the empty process can stay in every type of ambient); \trule{Par} gives to parallel composition of processes \PP\ and \Q\  the union of the sets of groups $\Gbar_1$ and $\Gbar_2$ obtained by typing \PP\ and \Q; rule \trule{Amb} checks whether a process \PP\ can be safely nested in an ambient \aname\ of type \G: if \PP\ is typed with  a set of types \Gbar\ we have to ensure that every type $\G_k$ in \Gbar\ can stay in an ambient of type \G; moreover, all  capability types collected in \de\ while typing \PP\ must be compatible with \G; since the scope of the capabilities is  the enclosing ambient, once the capabilities in \de\ have been checked to be admissible, \de\ is emptied; rule \trule{Cap}  verifies the correspondence between the type of a name used for capability synchronization and the capability prefix used with it and then adds the type to \de; rule \trule{Choice} gives to the choice between  \PP\ and \Q\  the union of the sets of groups $\Gbar_1$ and $\Gbar_2$ derived by typing \PP\ and \Q.

\begin{figure}[t]
	\centering
		\begin{tabular}{c}
\typerDe{\nul}{\G}{\emptyset}		\trule{Inact} \quad
	\typdG{\n}{\tname} 	\trule{Name}			
	\\
	\\
			\begin{prooftree}
		\typerD{\PP}{\Gbar}
		\justifies
		\typerD{ \restr{\cname}{\PP}}{\Gbar}
		\using	\trule{Restr}
		\end{prooftree}
		
\quad 		\begin{prooftree}
			\typerDe{\PP}{\Gbar_1}{\de_1}\quad 	\typerDe{\Q}{\Gbar_2}{\de_2}
		\justifies
			\typerDe{\PP \pa \Q}{\overline{\G_1,\G_2}}{\de_1 \cup\de_2}
		\using	\trule{Par}
		\end{prooftree}
			\quad
				\begin{prooftree}
		\typerD{\PP}{\Gbar}
		\justifies
		\typerD{ \repl{\PP}}{\Gbar}
		\using
			\trule{Repl}
		\end{prooftree}
		\\
		\\
		\begin{prooftree}	
	
\typd{\aname}{\gt} \quad \typerD{\PP}{\Gbar} \quad \G \in \SGe{\G_k},  \  \ \  \forall   \G_k \in \Gbar
\quad
\comp{\sy}{\G} \ \ \forall \ \ \sy \in \de
		\justifies
		\typerDe{\amb{\aname}{\PP}}{\G}{\emptyset}
		\using
			\trule{Amb}
		\end{prooftree}
			\\	
			\\
		\begin{prooftree}
	\typd{\cname}{\ch{\tname}} \quad \typerG{\ga, \n:\tname}{\PP}{\Gbar}
		\justifies
\typerD{\inG{\cname}{\n}.\PP}{\Gbar}
		\using
			\trule{Input}
		\end{prooftree}
		\quad
				\begin{prooftree}
	\typd{\cname}{\ch{\tname}} \quad \typerD{\PP}{\Gbar} \quad \typer{\m}{\tname}
		\justifies
\typerD{\outG{\cname}{\m}.\PP}{\Gbar}
		\using
			\trule{Out}
		\end{prooftree}
			\\
			\\	\begin{prooftree}
	\typd{\h}{\sy} \quad \capb \in \sy \quad \typerD{\PP}{\Gbar}
		\justifies
\typerDe{\capb \h.\PP}{\Gbar}{\de \cup \{\sy\}}
		\using
			\trule{Cap}
		\end{prooftree}
			\quad
			\begin{prooftree}
			\typerDe{\PP}{\Gbar_1}{\de_1} \quad 	\typerDe{\Q}{\Gbar_2}{\de_2}
		\justifies
			\typerDe{\PP + \Q}{\overline{\G_1,\G_2}}{\de_1 \cup \de_2}
		\using	\trule{Choice}
		\end{prooftree}
				\\
				\\
		\end{tabular}
	\caption{Typing rules.}
	\label{tab:Typing}
\end{figure}

We now show an example motivating the presence of the \CGn\ set of ambient groups that can be crossed.
Hydrophilic molecules are typically charge--polarized and capable of hydrogen bonding, thus enabling it to dissolve more quickly in water than in oil. Hydrophobic molecules instead tend to be non--polar and thus prefer other neutral molecules and non--polar solvents. As a consequence, hydrophobic molecules in water tend to cluster together forming micelles.  Hydrophobic molecules can cross cell membranes in a natural (and slow) way, even if there is no particular transporter on the membrane. On the contrary, hydrophilic molecules can cross membranes only with dedicated transporters (conveyers). We can model these \emph{crossing} properties with our type system. Namely, we can represent cells with or without conveyers as ambients of type \GE{CConv} and \GE{C} respectively; molecules can be of type \GE{Hphi} (hydrophilic) and \GE{Hpho} (hydrophobic). Finally, transporters have type \GE{Conv}. Molecules of types \GE{Hphi} and  \GE{Hpho} can stay in both  \GE{CConv} and \GE{C} cells but only \GE{Hpho} molecules can cross \GE{C} cells. The sets \SGn\ and \CGn\ associated to the types are given in Figure \ref{cellmol}.

	\begin{figure}[t]
	\centering
		\begin{tabular}{ccc}
		Group types \G & \SG
		 &   \CG \\
		\hline
		\\
		\GE{C}& \GE{Univ} & \GE{Univ}\\
	  \GE{CConv}& \GE{Univ} & \GE{Univ}\\
	  \GE{Conv}& \GE{CConv} & \GE{Univ}\\
		\GE{Hphi}& \GE{CConv}, \GE{C}  & \GE{CConv} \\
		\GE{Hpho}& \GE{CConv}, \GE{C}  & \GE{CConv}, \GE{C}\\
	\end{tabular}	
	\caption{Types for molecules and cells.}
	\label{cellmol}
\end{figure}

\noindent
Let
\[
\begin{array}{l}
\ga = cellC: \GE{CConv}, \ cell: \GE{C}, \ conv :   \GE{Conv}, \  \h': (\GE{Hphi},\GE{C})^\ee, \ \h'': (\GE{Hpho},\GE{C})^\ea,\\
 mol_1: \GE{Hphi}, \  mol_2: \GE{Hpho}, \ \h_1: (\{\GE{Hphi},\GE{Hpho}\},\GE{Conv})^\ea, \ \h_2: (\GE{Conv},\GE{CConv})^\ee, \ \\ \h_3: (\GE{Conv},\GE{CConv})^\ea,\ \h_4: (\{\GE{Hphi},\GE{Hpho}\},\GE{Conv})^\ee
 \end{array}
\]
\noindent
A cell with conveyors can be modeled as an ambient of type  \GE{CConv} with  nested conveyors and molecules:

\[
\amb{cellC}{\repl{\amb{conv}{\PP}} \pa \amb{mol_1}{\enter{\h}.\exit{\h_4}} \pa \amb{mol_2}{\exit{\h'}+\enter{\h}.\exit{\h_4}} \pa \expel{\h'}}
\]

\noindent
where $\PP= \accept{\h}.\enter{\h_1}.\exit{\h_2}.\enter{\h_3}.\expel{\h_4}$. Thus we model the conveyor as  first accepting molecules through \h \ then exiting the current cell through $\h_2$, entering a new cell through $\h_3$ \ and finally releasing it through $\h_4$.
Molecules of type \GE{Hpho} ($mol_1$) can enter inside the conveyor through \h \ and finally be expelled by it, after the transport, through $\h_4$; instead molecules of type \GE{Hpho} ($mol_2$) can also pass the membrane cell without the use of a conveyor (through $\h'$).

\section{Typed Operational semantics}\label{sec:TypedOpSem}

\begin{figure}[t]

	\centering
		\begin{tabular}{ll}
	\PP\pa\Q 	\stc \Q\pa \PP &
		(\PP\pa\Q )\pa \R 	\stc \PP\pa(\Q \pa \R) \\
\PP\pa\nul	\stc \PP   &
	 \restr{\n}{\nul} \stc \nul \\
\restr{\n}{\restr{\m}{\PP}} \stc	\restr{\m}{\restr{\n}{\PP}} &
\restr{\n}{\PP \pa \Q} 	\stc \PP \pa \restr{\n}{\Q}	if \n $\notin$ fn(\PP) \\
 \restr{\n}{\amb{\aname}{\PP}} 	\stc \amb{\aname}{\restr{\n}{\PP}}  &
\inG{\cname}{\m}.\PP	\stc \inG{\cname}{\n}.\subst{\PP}{\m}{\n}	if \n $\notin$ fn(\PP) \\
\restr{\m}{\PP } 	\stc \restr{\n}{\subst{\PP}{\m}{\n}}	if \n $\notin$ fn(\PP)		 &
	\repl{\nul} \stc \nul\\
	\repl{\PP} \stc \PP \pa \repl{\PP}
	
		\end{tabular}
	\caption{Structural congruence. 	\label{tab:strctcong}}

\end{figure}

\begin{figure}[t]
{\small	\centering
		\begin{tabular}{cl}	
		\red{\amb{\aname}{(\T + \enter{\h}.\PP) \pa \Q}\pa \amb{\bname}{(\T' + \accept{\h}.\R) \pa \SP}}{\amb{\bname}{\amb{\aname}{\PP \pa \Q} \pa \R \pa \SP}} & \trule{Red In}\\	
		\\
			\begin{prooftree}
				\typd{\bname}{\G_b}
				\justifies
		\red{\amb{\aname}{\amb{\bname}{(\T + \exit{\h}.\PP) \pa \Q} \pa (\T' + \expel{\h}.\R) \pa \SP}}{\amb{\bname}{\PP \pa \Q }\pa \amb{\aname}{\R \pa \SP}} \pa \warn{\G_b}
		\end{prooftree}& \trule{Red Out}
		\\
		\\
		\begin{prooftree}
\red{\PP}{\R\pa\warn{\G_i} }  \quad \typd{\aname}{\G_a}  \quad   \G_a \in \SGe{\G_i}
\justifies
	\red{ \amb{\aname}{\PP}}{\amb{\aname}{\R}}
\end{prooftree} & \trule{Red Amb WarnOK}
\\
\\
\begin{prooftree}
\red{\PP}{\R\pa\warn{\G_i} } \quad \typd{\aname}{\G_a} \quad  \G_a \notin \SGe{\G_i}
\justifies
	\red{ \amb{\aname}{\PP}}{ \exerror{\G_a}{\G_i}}
\end{prooftree} & \trule{Red Amb Warning}
\\	
		\\
		\begin{prooftree}
\red{\PP}{\Q}
\justifies
	\red{ \amb{\aname}{\PP}}{\amb{\aname}{\Q }}
\end{prooftree} & \trule{Red Amb }
		\\
		\\
	
	\begin{prooftree}
	\typd{\aname}{\G_a} \quad \typd{\R}{\Gbar_\R} \quad \typd{\SP}{\Gbar_\SP} \quad \forall \G_i \in (\Gbar_\R,\Gbar_\SP),\  \G_a \in \SGe{\G_i}
	\justifies
	\red{\amb{\aname}{(\T + \mergep{\h}.\PP) \pa \Q} \pa \amb{\bname}{(\T' + \mergem{\h}.\R) \pa \SP}}{\amb{\aname}{\PP \pa \Q \pa \R \pa \SP}} \end{prooftree}& \trule{Red Merge}
		\\		
	\\
		\begin{prooftree}
	\typd{\aname}{\G_a} \quad \typd{\R}{\Gbar_\R} \quad \typd{\SP}{\Gbar_\SP} \quad \exists \G_i \in (\Gbar_\R,\Gbar_\SP),\  \G_a \notin \SGe{\G_i}
	\justifies
	\red{\amb{\aname}{(\T + \mergep{\h}.\PP) \pa \Q} \pa \amb{\bname}{(\T' + \mergem{\h}.\R) \pa \SP}}{\merror{\G_a}{\G_i}}	 \end{prooftree} & \trule{Red Merge Error}
	\\
	\\
		\red{(\T + \inE{\loc}{\cname}{\m}.\PP) \pa (\T' + \outE{\loc}{\cname}{\n}.\Q)}{\subst{\PP}{\m}{\n} \pa \Q } & \trule{Red Local}\\	
	\\
		\red{(\T + \outE{\ptc}{\cname}{\n}.\PP) \pa \amb{\aname}{(\T' + \inE{\ctp}{\cname}{\m}.\Q)}\pa \R}{\PP \pa \amb{\aname}{\subst{\Q}{\m}{\n} \pa \R } }& \trule{Red Parent Output}\\	
	\\
		\red{\amb{\aname}{(\T + \outE{\ctp}{\cname}{\n}.\PP)\pa \R} \pa (\T' + \inE{\ptc}{\cname}{\n}.\Q) }{\amb{\aname}{\R \pa\PP} \pa \subst{\Q}{\m}{\n} }& \trule{Red Parent Input}\\	
	\\
		\red{\amb{\aname}{(\T + \outE{\sts}{\cname}{\n}.\PP)\pa \R} \pa \amb{\bname}{(\T' + \inE{\sts}{\cname}{\m}.\Q) \pa \SP}}{\amb{\aname}{\R \pa\PP} \pa \amb{\bname}{\subst{\Q}{\m}{\n}\pa \SP }}& \trule{Red Sibling}
\\
\\
		\end{tabular}
			\begin{tabular}{l}
	\begin{prooftree}
\red{\PP}{\Q}	 \justifies \red{ \restr{\n}{\PP}}{ \restr{\n}{\Q}} \using \trule{Red Res} \end{prooftree}    \quad

	\begin{prooftree}\red{\PP}{\Q}	\justifies \red{\PP \pa \R}{ \Q \pa \R}\using \trule{Red Par} \end{prooftree}  \quad
 	\begin{prooftree}\congr{\PP}{\PP'},  \red{\PP}{\Q},  \congr{\Q}{\Q'}	\justifies \red{\PP'}{\Q'} \using \trule{Red $\equiv$}\end{prooftree}  \\
\\
		\end{tabular}
	\caption{Operational Semantics}
	\label{tab:OperationalSemantics}}
\end{figure}

In this section we extend  the semantics of BioAmbients  by adding rules which rise errors as a consequence of  undesired behaviour.
The structural congruence  of BioAmbients remains unchanged, we recall it in  Figure~\ref{tab:strctcong}.
Rules for ambients movements and communications model \emph{reactions} which may happen when two complementary prefixes on the same name \n\ occur in parallel.
Safety of communications and \ea\ capabilities can be statically checked by typing rules: \ea\ capabilities are ensured to be well formed, i.e. they cannot move an ambient \amb{\aname}{\ldots} of type \G\ in an ambient \amb{\bname}{\ldots } of type \G'\ if $\G' \notin \SG$ (see Definiton \ref{def:wf}).
On the other hand, a static control of \ee\ and \mm\ capabilities would require too many constraints in the definition of group types: we should check the relation between the  group types involved in all possible \ee\ and \mm\ interactions; as a consequence the type system would be very restrictive   discarding also  safe reductions just because of the presence of  potentially unsafe capability prefixes in a choice. For this reason we check \ee\ and \mm\  reductions at run-time, signalling errors when they arise.
The reduction rules are in Figure \ref{tab:OperationalSemantics}.

Rule \trule{RedIn} reduces the synchronization (thorough a name \h)  of \enter{\h}/\accept{\h} capability to the entrance of an ambient \amb{\aname}{\ldots} in an ambient \amb{\bname}{\ldots}. As explained above, if this rule is applied to a well typed process after the reduction the  nesting of  ambients is safe.
Rule \trule{RedOut} reduces the synchronization of \exit{\h}/\expel{\h} prefixes  to the exit of an ambient \amb{\bname}{\ldots} out of an ambient \amb{\aname}{\ldots}. We put a warning  \warn{\G_b} in parallel with the new sibling ambients, since we do not know in which ambient \amb{\bname}{\ldots} will arrive once exited  from \amb{\aname}{\ldots}: e.g. \amb{\bname}{\ldots}\  could be nested in an ambient where it cannot stay. Two  rules  model the reduction of warned ambients inside another ambient:  \trule{Red Amb WarnOK} reduces the warning  parallel to a safe one if the exit did not produced an unsafe nesting; on the contrary, in case of unsafe nesting \trule{Red Amb Warning} generates an error; finally rule \trule{Red Amb} models reduction inside an ambient when there are no warnings.

Rule \trule{Merge} reduces the synchronization of \mergep{\h}/\mergem{\h}  prefixes to the fusion of two sibling ambients into a single one: the \mergem{} prefix "brings" all the processes in parallel with the prefixed one into the sibling ambient.   We cannot  check statically which processes will be in parallel with the  prefixed one when the reduction rule is applied: we perform this check at runtime raising an error in case of unsafe nesting due to the merging of two sibling ambients (rule \trule{Red Merge Warning}).
Concerning communications, rules are unchanged w.r.t. \cite{BA},  they model names substitutions due to communications between processes  located in same ambient (\trule{Red Local}), in parent-child ambients (\trule{Red Parent Output}, \trule{Red Parent Input}) and in sibling ambients \trule{Red Sibling}.

Note that in the rules of our operational semantics there are no checks on the sets \CGn\ since capability types should be well formed (thus satisfying the conditions for \CGn\ sets).

\noindent
Let \error{\G_i}{\G_j} range over \exerror{\G_i}{\G_j} and \merror{\G_i}{\G_j}. A well typed process either reduces to another well typed process or generates an error.
\begin{theorem}
If \ \typerD{\PP}{\Gbar}    then
\begin{itemize}
	\item either $\red{\PP}{\PP'}$ or \red{\PP}{\PP' \pa \warn{\G}} and $\exists \ \ga', \Gbar', \de'$ such that \typerDe{\PP'}{\Gbar'}{\de'}
	\item or  $\exists \ \G_i, \ \G_j \mbox{ such that  } \red{\PP}{\error{\G_i}{\G_j}}$
\end{itemize}
\noindent
\textbf{Proof.} By induction on the definition of $\rightarrow$.
\end{theorem}

Note that our semantics does not reduce the warnings \warn{\G} at top level. While they do not affect the system evolution, they could be useful in a compositional setting. In particular, if the entire process should be nested, at some point, into another ambient, the warnings keep the conditions on the admissible ambients (without the need to recompute the whole type of the system).

	\section{Motivating Examples}\label{sec:Examples}
	In this section we provide a couple of simple but motivating examples. In the following we will use \aname\ instead of \amb{\aname}{\nul} and we assume \SG=\CG\ whenever \CGn\ is not explicitly represented.

	\subsection{Blood transfusion}
This example has been inspired by~\cite{TR09}.
	A blood type is a classification of blood based on the presence or absence of inherited antigenic substances on the surface of red blood cells: these antigens are the A antigen and the B antigen. Blood type A contains only A antigens, blood type B contains only B antigens, blood type AB contains both and the blood type O contains none of them.

The immune system will produce antibodies that can specifically bind to a blood group antigen that is not recognized as self: individuals of blood type A have Anti-B antibodies, individuals of blood type B have Anti-A antibodies, individuals of blood type O have both Anti-A and Anti-B antibodies, and individuals of blood type AB have none of them. These antibodies can bind to the antigens on the surface of the transfused red blood cells, often leading to the destruction of the cell: for this reason, it is vital that compatible blood is selected for transfusions.

Another antigen that refines the classification of blood types is the RhD antigen: if this antigen is present, the blood type is called positive, else it is called negative. Unlike the ABO blood classification, the RhD antigen is immunogenic, meaning that a person who is RhD negative is very likely to produce Anti-RhD antibodies when exposed to the RhD antigen, but it is also common for RhD-negative individuals not to have Anti-RhD antibodies.
We model blood transfusion  as a system consisting of a set of closed tissues. Tissues contain blood cells and antibodies according to the classification described above, then they can join each other  performing a transfusion of different blood types. We model a red blood cell as an ambient whose type represents the blood type; thus, the  groups representing blood types are: \GE{A+}, \GE{A-}, \GE{B+}, \GE{B-}, \GE{AB+}, \GE{AB-}, \GE{O+}, \GE{O-}. We represent    A,B, RhD antigen and  Anti-A, Anti-B and Anti-RhD antibodies as ambients of type \GE{a},	\GE{b}, \GE{r},	 \GE{\overline{a}}, \GE{\overline{b}},	\GE{\overline{r}}  respectively. The sets \SG\ associated to the different blood types are given in Figure \ref{tab:TypeForBloodGroups}. Finally, we model a tissue (which contains the red cells) as an ambient of type $\G_i \in \{\GE{A+}, \GE{A-}, \GE{B+}, \GE{B-}, \GE{AB+}, \GE{AB-}, \GE{O+}, \GE{O-}\}$.

	\begin{figure}[t]
	\centering
		\begin{tabular}{cc}
		Group types  & \SG\\
		of basic elements & \\
		\hline
		\\
		\SGe{\GE{a}}& \GE{A-}, \GE{AB-}, \GE{A+}, \GE{AB+}  \\
		\SGe{\GE{b}} & \GE{B-}, \GE{AB-}, \GE{B+}, \GE{AB+} \\
		\SGe{\GE{r}} & \GE{A+},\GE{B+},  \GE{AB+}, \GE{O+} \\
		\SGe{\GE{\overline{a}} }& \GE{B+}, \GE{B-}, \GE{O+} ,\GE{O-}   \\
		\SGe{\GE{\overline{b}} }& \GE{A+}, \GE{A-}, \GE{O+}, \GE{O-} \\
		\SGe{\GE{\overline{r}} }& \GE{A-},\GE{B-}, \GE{AB-}, \GE{O-}   \\
	\end{tabular}	
	\caption{Types for blood groups.}
	\label{tab:TypeForBloodGroups}
\end{figure}
We model blood transfusion as the reduction between two tissues having complementary merge capabilities (\mergem{}\ for the donor, \mergep{}\ for the receiver). For instance let us consider a tissue	$\tname_1$  represented by an ambient of type  \GE{A+} and two potential donors $\tname_2$ and $\tname_3$ of types \GE{B+} and \GE{O+} respectively:
\[
\PP= \amb{\tname_1}{ \repl{(\mergepn \h_1 +  \mergepn \h_2 + \ldots  \mergepn \h_n )}\pa a_1 \pa \overline{b_1} \pa r_1} \pa
	 \amb{\tname_2}{\mergem \h_1 \pa b_1 \pa r_2} \pa
	  \amb{\tname_3}{\mergem \h_2  \pa r_3}
\]

\noindent
\PP\  is well typed with
\[
\begin{array}{l}
\ga =\tname_1:\GE{A+},\ \tname_2:\GE{B+},\ \tname_3:\GE{O+},\ a_1: \GE{a},  \  \overline{b_1} : \GE{ \overline{b}}, \   r_1: \GE{r}, \  r_2: \GE{r}, \ r_3: \GE{r},  \  b_1: \GE{b},\ \\
\h_1: (\GE{A+},\GE{B+})^\mm,  \h_2: (\GE{A+},\GE{O+})^\mm, \ldots
\end{array}
\]

\noindent
Thus, the tissue $\tname_1$ can potentially receive blood from many donors ($\mergepn \h_1 +  \mergepn \h_2 + \ldots  \mergepn \h_n $), and, because for example of some human error, may also receive blood which is not compatible to its own. Let us consider two possible reductions. The first one:
\[
\red{\amb{\tname_1}{\repl{( \mergepn \h_1 +  \mergepn \h_2 + \ldots  \mergepn \h_n)}  \pa a_1 \pa \overline{b_1} \pa r_1} \pa \amb{\tname_2}{\mergem h_1 \pa b_1 \pa r_2}}{\merror{\GE{A+}}{\GE{b}}}
\]
\noindent
results in an error because of a  wrong transfusion causing the presence of an antigen of type \GE{b} in a tissue of type \GE{A+}. The second one:
\[
\begin{array}{l}
	\amb{\tname_1}{\repl{( \mergepn \h_1 +  \mergepn \h_2 + \ldots  \mergepn \h_n)}  \pa a_1 \pa \overline{b_1} \pa r_1} \pa \amb{\tname_3}{\mergem h_2  \pa r_3} \rightarrow \\
	\amb{\tname_1}{\repl{(\mergepn \h_1 +  \mergepn \h_2 + \ldots  \mergepn \h_n )} \pa a_1 \pa \overline{b_1} \pa r_1 \pa r_3}
\end{array}
\]
\noindent
models a transfusion between compatible blood types, namely $A+$ and $O+$.

\subsection{Bacteriophage viruses}
\begin{figure}
\centering
\includegraphics[width=6cm]{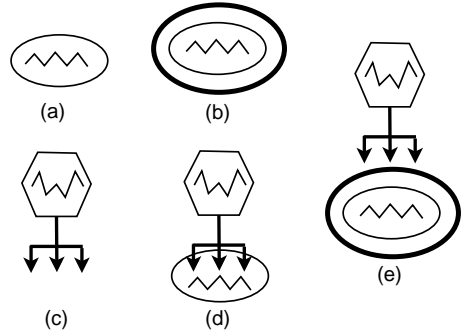}
\caption{\label{batteriofago} A bacterium (a) could be represented as a membrane containing the bacterium's DNA. A resistent (coated) spore (b) is represented as a bacterium surrounded by its coat. A bacteriophage (c) is depicted with the outer capsid, containing the genetic material, and the hypodermic syringe, used to inject its genetic material into the bacteria cells (d). They cannot inject coated cells (e).}
\end{figure}

In this section we use our system to model the interaction  between bacteria and bacteriophage viruses (see Figure~\ref{batteriofago}).

We  assume that a bacterium consists
of a cellular membrane containing its DNA. The sporulation
mechanism allows producing inactive and very resistant bacteria
forms, called spores which are surrounded by a membrane (coat) protecting them from virus attacks. A spore can germinate and then produce a new bacterium.
 A bacterium can safely stay in ambients containing viruses if it is protected by its coat.
 The types involved in this model are: \GE{EnvOk},  \GE{EnvVirus} are the types of environments respectively virus-free and  virus-friendly; \GE{Bact},\GE{Coat} are the types of the bacteria and the protecting membrane. \GE{Vir} is the type of viruses. The corresponding (relevant) \SGn\ groups  are shown in Figure \ref{ex:bact}.

\begin{figure}
\centering
		\begin{tabular}{cc}
		Group types  & \SG\\
		of basic elements & \\
		\hline
		\\
	\GE{Bact}& \GE{EnvOk} \\
		\GE{Coat} & \GE{EnvOk}, \GE{EnvVirus} \\
	\GE{Vir} & \GE{EnvVirus} \\
	\end{tabular}		\caption{Bacteria-Viruses Example: Types}
	\label{ex:bact}
\end{figure}
\noindent
Now let us consider a bacteria $\bname_2$ surrounded by a coat  $\bname_1$ (we omit the description of the DNA inside the bacterium):
\[
\PP = \amb{\bname_1}{\amb{\bname_2}{\exit{\h}\pa \enter{\h_2}} \pa   \repl{(\expel{\h} +\enter{\h_1}+\enter{\h_2})}}
\]
\noindent
\PP \ is well typed with:
\[
\begin{array}{l}
\ga = \h:(\mbox{\GE{Bact},\GE{Coat}})^\ee, \ \h_1:(\mbox{\GE{Coat},\GE{EnvOK}})^\ea, \ \h_2:(\mbox{\GE{Coat},\GE{EnVirus}})^\ea,  \\  \bname_1:\GE{Coat}, \ \bname_2:\GE{Bact}, \ \aname_1:\GE{EnvVirus}\aname_2:\GE{EnvOk}
\end{array}
\]
\noindent
We represent the chance of the bacterium to get rid of the coat as an exit/expel capability through the name \h\ which allows the bacterium to germinate (exiting from its protecting membrane). The coat (containing the bacterium) can move in every environment, while the bacterium can only enter \GE{EnvOk} environments; this is modeled by the use of suitable \ea\ capabilities.
We now put the two environments $\aname_1$ (allowing viruses) and $\aname_2$ (virus-free) in parallel with the bacterium $\bname_1$.
There are three possible behaviour (in the following we put labels on transitions for the sake of readability):

\begin{enumerate}
	\item The bacterium gets rid of the coat and then can enter only in $\aname_2$:

\begin{tabular}{c}
\amb{\bname_1}{\amb{\bname_2}{\exit{\h}\pa \enter{\h_2}} \pa   \repl{(\expel{\h} +\enter{\h_1}+\enter{\h_1})}} \pa \amb{\aname_1}{\repl{\accept{\h_1}}}\pa \amb{\aname_2}{\repl{\accept{\h_2}}}\\
 $\stackrel{\ee(\h)}{\rightarrow} $\\
 \amb{\bname_1}{ \repl{(\expel{\h} +\enter{\h_1}+\enter{\h_1})}} \pa \amb{\bname_2}{ \enter{\h_2}}\pa \warn{\amb{\aname_1}{\repl{\accept{\h_1}}}}\pa \amb{\aname_2}{\repl{\accept{\h_2}}}\\
   $\stackrel{\ea(\h_2)}{\rightarrow} $\\
   \amb{\bname_1}{ \repl{(\expel{\h} +\enter{\h_1}+\enter{\h_1})}} \pa \amb{\aname_1}{\repl{\accept{\h_1}}}\pa \amb{\aname_2}{\amb{\bname_2}{} \pa \repl{\accept{\h_2}}}
\end{tabular}

\item The coat can move in the ambient $\aname_1$ and then expel the bacterium in this hostile environment (thus generating an error):

\begin{tabular}{c}
 \amb{\bname_1}{\amb{\bname_2}{\exit{\h}\pa \enter{\h_2}} \pa   \repl{(\expel{\h} +\enter{\h_1}+\enter{\h_1})}}
  \pa \amb{\aname_1}{\repl{\accept{\h_1}}}\pa \amb{\aname_2}{\repl{\accept{\h_2}}}\\
 $\stackrel{\ea(\h_1)}{\rightarrow} $\\
 \amb{\aname_1}{\amb{\bname_1}{\amb{\bname_2}{\exit{\h}\pa \enter{\h_2}} \pa   \repl{(\expel{\h} +\enter{\h_1}+\enter{\h_2})}} \pa\repl{\accept{\h_1}}}\pa \amb{\aname_2}{\repl{\accept{\h_2}}}\\
 $\stackrel{\ee(\h)}{\rightarrow} $\\
  \amb{\aname_1}{\amb{\bname_1}{   \repl{(\expel{\h} +\enter{\h_1}+\enter{\h_2})}} \pa \warn{\amb{\bname_2}{\exit{\h}\pa \enter{\h_2}} }\pa \repl{\accept{\h_1}}}\pa \amb{\aname_2}{\repl{\accept{\h_2}}} \\
   $\rightarrow $\\
   \exerror{\GE{EnvVirus}}{\GE{Bact}}
\end{tabular}

\item The coat can move in the ambient $\aname_2$ and then expel the bacterium:

\begin{tabular}{c}
 \amb{\bname_1}{\amb{\bname_2}{\exit{\h}\pa \enter{\h_2}} \pa   \repl{(\expel{\h} +\enter{\h_1}+\enter{\h_1})}}
  \pa \amb{\aname_1}{\repl{\accept{\h_1}}}\pa \amb{\aname_2}{\repl{\accept{\h_2}}}\\
 $\stackrel{\ea(\h_2)}{\rightarrow} $\\
 \amb{\aname_2}{\amb{\bname_1}{\amb{\bname_2}{\exit{\h}\pa \enter{\h_2}} \pa   \repl{(\expel{\h} +\enter{\h_1}+\enter{\h_2})}} \pa\repl{\accept{\h_2}}}\pa \amb{\aname_1}{\repl{\accept{\h_1}}}\\
 $\stackrel{\ee(\h)}{\rightarrow} $\\
  \amb{\aname_2}{\amb{\bname_1}{   \repl{(\expel{\h} +\enter{\h_1}+\enter{\h_2})}} \pa \warn{\amb{\bname_2}{\exit{\h}\pa \enter{\h_2}} }\pa \repl{\accept{\h_2}}}\pa \amb{\aname_1}{\repl{\accept{\h_1}}} \\
   $\rightarrow $\\
  \amb{\bname_1}{ \repl{(\expel{\h} +\enter{\h_1}+\enter{\h_1})}} \pa \amb{\aname_1}{\repl{\accept{\h_1}}}\pa \amb{\aname_2}{\amb{\bname_2}{\enter{\h_2}} \pa \repl{\accept{\h_2}}}
\end{tabular}
\end{enumerate}

\section{Conclusions}\label{sec:Conc}

The most common approach of biologists to describe biological systems is based on the use of
deterministic mathematical means (like, e.g., ODE), and makes it possible to abstractly reason on the behaviour of
biological systems and to perform a quantitative \emph{in silico} investigation. This kind of modelling,
however, becomes more and more difficult, both in the specification phase and in the analysis processes, when
the complexity of the biological systems taken into consideration increases. This has probably been one of the main
motivations for the application of Computer Science formalisms to
the description of biological systems~\cite{RegSha02}. Other
motivations can also be found in the fact that the use of formal methods from
Computer Science permits the application of analysis techniques that
are practically unknown to biologists, such as, for example, static analysis and
model checking.

Different formalisms have either been applied to (or have been inspired
from) biological systems. The most notable are automata-based models
\cite{AluBelKumMin01,MatDoiNagMiy00}, rewrite systems
\cite{DanLan03a,Pau02}, and process calculi
\cite{RegSha02,RegSha04,BA,Car04,PriQua05}. Automata-based models have the
advantage of allowing the direct use of many verification tools, such as,
for example, model checkers. On the other side, models
based on rewrite systems describe biological systems with a notation
that can be easily understood by biologists. However, automata-like
models and rewrite systems are not compositional. The possibility to study in a componentwise way the
behaviour of a system  is, in general, naturally ensured by process
calculi, included those commonly used to describe biological
systems.

In this paper we have laid the foundations for a type system for the BioAmbients calculus suitable to guarantee compatible compartments nesting (due to some intrinsical biological properties). In this framework, the correctness of the \emph{enter}/\emph{accept} capabilities can be checked statically, while the \emph{merge} capability and the \emph{exit}/\emph{expel} capabilities could cause the movement of an ambient of type $G$ within an ambient of type $G'$ and dynamically rise an error. We used our type discipline to model how incompatible blood transfusion could cause the system to rise an error, or to represent the movement of bacteria spore into friendly environments where they can germinate and restart their activity. 

\vspace{0.5cm}
\noindent
\textbf{Acknowledgments} We would like to warmly thank Mariangiola Dezani-Ciancaglini who encouraged us
to write this paper and gave us crucial suggestions.

\bibliographystyle{plain} 
\bibliography{biblio}
\end{document}